Original Paper

# Convolutional Neural Network–Based Automatic Classification of Colorectal and Prostate Tumor Biopsies Using Multispectral Imagery: System Development Study

Remy Peyret[1], PhD; Duaa alSaeed[2], PhD; Fouad Khelifi[1], PhD; Nadia Al-Ghreimil[2], PhD; Heyam Al-Baity[2], PhD; Ahmed Bouridane[1], PhD

[1]Northumbria University at Newcastle, Newcastle, United Kingdom
[2]College of Computer and Information Sciences, King Saud University, Riyadh, Saudi Arabia

**Corresponding Author:**
Duaa alSaeed, PhD
College of Computer and Information Sciences
King Saud University
King Abdullah Road
Riyadh, 11451
Saudi Arabia
Phone: 966 555442477
Email: dalsaeed@ksu.edu.sa

## Abstract

**Background:** Colorectal and prostate cancers are the most common types of cancer in men worldwide. To diagnose colorectal and prostate cancer, a pathologist performs a histological analysis on needle biopsy samples. This manual process is time-consuming and error-prone, resulting in high intra- and interobserver variability, which affects diagnosis reliability.

**Objective:** This study aims to develop an automatic computerized system for diagnosing colorectal and prostate tumors by using images of biopsy samples to reduce time and diagnosis error rates associated with human analysis.

**Methods:** In this study, we proposed a convolutional neural network (CNN) model for classifying colorectal and prostate tumors from multispectral images of biopsy samples. The key idea was to remove the last block of the convolutional layers and halve the number of filters per layer.

**Results:** Our results showed excellent performance, with an average test accuracy of 99.8% and 99.5% for the prostate and colorectal data sets, respectively. The system showed excellent performance when compared with pretrained CNNs and other classification methods, as it avoids the preprocessing phase while using a single CNN model for the whole classification task. Overall, the proposed CNN architecture was globally the best-performing system for classifying colorectal and prostate tumor images.

**Conclusions:** The proposed CNN architecture was detailed and compared with previously trained network models used as feature extractors. These CNNs were also compared with other classification techniques. As opposed to pretrained CNNs and other classification approaches, the proposed CNN yielded excellent results. The computational complexity of the CNNs was also investigated, and it was shown that the proposed CNN is better at classifying images than pretrained networks because it does not require preprocessing. Thus, the overall analysis was that the proposed CNN architecture was globally the best-performing system for classifying colorectal and prostate tumor images.

## Introduction

**Background**

According to the World Health Organization 2014 report, 14 million new cases of cancer were diagnosed in 2012, and the disease caused 8 million people to die in the same period [1]. Colorectal cancer is the third most common cancer globally, whereas prostate cancer is the second most common cancer among men, accounting for 9.7% and 7.9% of all cancers in both sexes, respectively [1]. Both colorectal and prostate tissues are glandular and therefore have a similar histological appearance.

For prostate cancer diagnosis, the European Association of Urology guidelines [2] recommend the performance of a histological analysis on a sample taken from a needle biopsy by a pathologist who decides the grade and stage of cancer or the type of tumor based on their experience and expertise. However, this process is time consuming and it also results in a high intra- and interobserver variability [3,4], which affects diagnosis reliability. In December 1999, a study [5] of more than 6000 patients conducted by Johns Hopkins researchers found that up to 2 out of every 100 people who came to larger medical centers for treatment were given an incorrect diagnosis after histological analysis. These results suggest that second-opinion pathology examinations not only prevent errors but also save lives and money. Consequently, there is an increasing interest among pathology experts in the use of machine vision (or computational diagnosis tools) to reduce diagnosis error rates by lowering the fallible aspect of human image interpretation.

Computer-aided diagnosis can assist pathologists in reducing the human analysis time, improving efficiency, and acting as a second opinion [6-8]. Adding computer-based quantitative analysis to human qualitative interpretation could significantly reduce the intra- and interobserver variability revealed in [4]. The main objective of this study is to develop an automatic computerized system for the diagnosis of colorectal and prostate tumors using images of biopsy samples.

Numerous investigations concerning prostate or colorectal tumor classification have been carried out [9,10]. However, most use color spaces limited to gray-scale or red, green, blue (RGB) images. In the last decade, many studies have used multispectral images [11-18], which are acquired using a more precise sampling of the light spectrum. This approach aims to better capture the spectrum of the reflected light coming from the observed sample, offering more discriminative information. Lasch et al [19] suggested that multispectral imagery can improve histopathological analysis by capturing patterns that are invisible to the human vision system and standard RGB imaging. Multispectral imaging studies have shown promising results and often outperformed systems using traditional gray-scale or RGB images [9,10]. However, multispectral images contain a large amount of data, making them more difficult to process because of increased execution time and problems caused by the curse of dimensionality [13].

Since the emergence of graphic processing units (GPUs) with sufficient processing power to train Convolutional neural networks (CNNs) in 2011, these models have seen a growing interest in image classification. Several models have been developed and tested on the ImageNet data set. As an example, the AlexNet architecture was developed in 2012 [20] and won several international competitions, including the ImageNet competition. GoogLeNet [21], a 22 layers deep network, won the ImageNet competition of 2014. He et al [22] deepened the networks even more with ResNet and won the best paper in 2015 at the Conference on Computer Vision and Pattern Recognition. To reduce training times, they developed a framework in which layers are formulated as a residual function with reference to the layer input, as opposed to the unreferenced learning functions previously used. The residual network comprised 152 layers. In 2016, Google DeepMind used a mix of supervised deep learning and reinforcement learning (ie, deep reinforcement learning) to create a system capable of learning how to play the game of *Go* [23]. This program, called AlphaGo, achieved a 99.8% winning rate against other Go programs and defeated the human European Go champion by 5 games to 0. In 2017, they created AlphaGo Zero [24], which outperformed the original AlphaGo in terms of performance and learning time without using any human knowledge. CNNs seem particularly adapted to the problem of microscopic images of tumor classification. A previous study [25] applied CNNs to microscopic images of colorectal cancer and found a promising accuracy of 99.1%. However, in this study, images were preprocessed using an active contour model before being fed to the CNN model. This operation requires the intervention of a pathologist to select the region of interest from the segmented image. Otherwise, this step can be replaced by another supervised learning model, which requires more training and thus dramatically increases the processing time. This study proposes a model that does not require a preprocessing phase and uses a single CNN model for the entire classification task using multispectral images.

Deep learning is a branch of machine learning that attempts to mimic the thinking process. To process data, information is passed through a network consisting of different layers, where each layer serves as input to the following layer. The first layer of a network is referred to as the input layer, whereas the last layer is the output layer. All the layers in between are called hidden layers. Typically, a layer is a simple algorithm that consists of an activation function. This field of machine learning is now very active, and the research community is focused on solving practical applications using modern deep learning. This study aims to apply the deep learning framework to the problem at hand.

**Objective**

The primary objective of this study is to develop a computerized automatic system for the diagnosis of colorectal and prostate tumors using images of biopsy samples to reduce time and diagnosis error rates associated with human analysis. To achieve this, we propose a CNN model for the classification of colorectal and prostate tumors from multispectral images of biopsy samples. The key idea is based on removing the last block of the convolutional layers and halving the number of filters per layer.

This paper is organized as follows: we first describe the principles of deep neural networks. The second section discusses the proposed method, whereas the data sets of multispectral tumor images are described in the third section. In the fourth section, the experiments carried out to validate the approach are detailed, and finally their results are presented and analyzed.

## Feedforward Neural Networks

### Overview

Feedforward neural networks, also called multilayer perceptrons (MLPs), are the basis of deep learning models. They aim to approximate the function f:~x!y, where ~x is an input feature vector and *y* is its corresponding class. The network builds a mapping ~y=f$_{(~x;)}$ by learning the parameters that provide the best approximation function to *f*. In this type of network, information moves from the input to the output through intermediate layers with no feedback connections. The number of layers is called the network depth. Each layer consists of a vector of functions or units that act in parallel, and the dimension of this vector is the width of the layer. Therefore, many hyperparameters need to be chosen when designing a neural network model, including its architecture, that is, the number of layers and units per layer.

A hidden layer computes an affine transformation of its input and then applies a nonlinear function *g*. This is defined by h=g$_{(W~x+b)}$, where *h* is the output of the hidden layer, *W* is the weight of the affine transformation, and *b* is the bias. *W* and *b* are the parameters learned when training the model.

The function chosen for each unit is called the activation function and is inspired by the behavior of biological neurons. The most widely used activation function is the rectified linear unit (ReLU), defined by g$_{(z)}$=max$_{(0, z)}$. Many other options are available, and the research on activation function is still a very active field. However, the ReLU has proven to perform well and is the default choice for activation functions.

Network training is performed using gradient descent. The main difference from other models is that the nonlinearity of neural networks causes the loss function to be nonconvex. Unlike convex optimization used with support vector machines or deep reinforcement learning, there is no guarantee of global convergence of a gradient descent applied to a nonconvex loss function. Consequently, the learning process is sensitive to the initial values of weights and biases. To apply gradient-based learning, a cost function must be chosen. The problem at hand in this study defines a conditional distribution $p(y/x; \theta)$ and the maximum likelihood principle is well adapted for it [26]. As a result, the cross-entropy between the training data and the model's prediction, which is equivalent to the negative log-likelihood, is used as the cost function. It enables the model to estimate the conditional probability of the classes if the input is known. The cost function model is as follows:

$$J(\theta) = -\mathbb{E}_{X,Y \sim \hat{p}_{data}} \log p_{model}(\mathbf{y}|\mathbf{x})$$

where $\hat{p}_{data}$ is the distribution of the training data and p$_{model}$ is the model distribution and the set of parameters for which the cost function is calculated. Consequently, the specific form of the cost function changes depending on the form of the log p$_{model}$.

### Back-Propagation

During training, the gradient of the cost function $\Delta_\theta J(\theta)$ is computed using a back-propagation algorithm [27-29] to allow information to flow backward through the network and compute the error made on each network weight. A gradient descent was then used to minimize the cost function. Learning was subsequently performed by updating the weights of the units. This procedure is described in the algorithm shown in Figure 1.

Training a neural network consists of applying a series of forwarding propagations—the network output is generated from the data through the network, and back-propagations compute the error at each unit. Each of these forward propagation and back-propagation combinations is called a pass. A pass of all the training examples is performed to compute the gradient used for the gradient-descent algorithm. A pass of every training example is called an epoch. At the end of each epoch, the network weights are updated using a learning rate hyperparameter, which is multiplied by the gradient calculated with back-propagation.

The learning rate is one of the most important hyperparameters for tuning in a neural network, as it controls the effective capacity of the network [26]. Therefore, it needs to be carefully optimized. If the learning rate is too large, the gradient descent can have the opposite of the desired effect, and training accuracy can decrease [30]. However, when it is too small, the training is slower, and sometimes the training accuracy can stay permanently small [30]. The number of epochs is also a hyperparameter that can be tuned ahead of the training.

**Figure 1.** Back-propagation algorithm.

```
Algorithm   Back-propagation algorithm for a L-layer network with
            weights θ^(l) and a training set {(x_1, y_1), ..., (x_m, y_m)}.
 1  for l ← 1 to L do
 2      θ^(l) = small random value ;      // Initialise network weights for
                                          each layer
 3  end
 4  foreach epoch do
 5      for l ← 1 to L do
 6          Δ^(l) = 0 ;                   // Initialise gradient matrices
 7      end
        // For each training example
 8      foreach (x_i, y_i) ∈ {(x_1, y_1), ..., (x_m, y_m)} do
            // Forward propagation
 9          w^(1) ← x_i;
10          for l ← 2 to L do
11              w^(l) ⇐ g(θ^(l-1) w^(l-1)) ;    // For each layer of the
                                                 network
12          end
            // Back-propagation
13          δ^(L) ← w^(L) − y_i ;               // Compute the error at the output
                                                 layer
14          for l ← L − 1 to 2 do
15              δ^(l) ⇐ ((θ^(l))^T δ^(l)) .* w^(l) .* (1 − w^(l)) ;     // Compute the
                    error of each unit at the hidden layers
16              Δ^(l) ← Δ^(l) + δ^(l) (w^(l))^T ;  // Update the matrix Δ for
                    each layer
17          end
18      end
        // Gradient-descent: Update weights using learning rate
            η and gradient (1/m)Δ
19      for l ← 1 to L do
20          θ^(l) ← θ^(l) − η (1/m) Δ^(l)
21      end
22  end
23  return θ^(1), ..., θ^(L);
```

## Methods

### Overview

As previously mentioned, the research community is now focusing on solving practical applications using deep learning approaches. Our proposed solution to the problem of diagnosing colorectal and prostate cancer is to apply a deep learning framework.

CNNs [27,31] are a type of neural network that specialize in data with a grid-like topology. They are particularly adapted for image processing. Similar to conventional neural networks, they consist of units with weights and biases that are learned during training. However, with the assumption of the data topology, it is possible to add some properties to the architecture to reduce the number of parameters to learn and improve the network implementation efficiency. These key ideas are local connections, shared weights, pooling, and the use of many layers [32].

The CNN units are arranged in three dimensions in each layer of the network: width, height, and depth of the activation volume. As depicted in Figure 2, a total of 3 different types of layers are usually stacked to form the full CNN architecture: convolutional layer, pooling layer, and fully connected layer. Fully connected layers are layers of a traditional MLP, as described in the section *Feedforward Neural Networks*.

**Figure 2.** Convolutional neural network architecture.

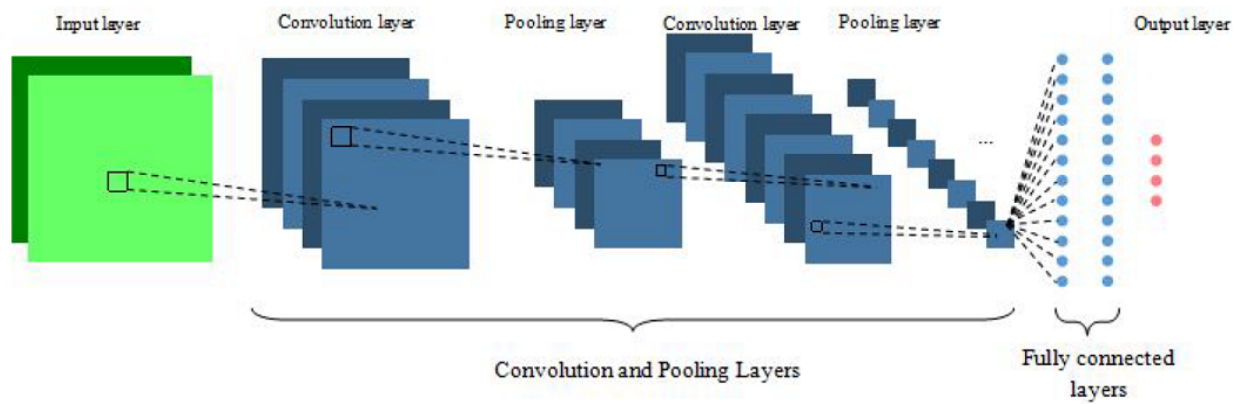

## Convolutional Layer

The convolutional layer is the core layer of a CNN. The basic idea is that instead of connecting a unit to every unit of the previous layer, it is only connected to a local region of the previous layer. The spatial extent of this connection is called the receptive field of the unit or filter size. This is a hyperparameter of the model. The filter size along the depth axis is the same as that of the previous layer. This shows an asymmetry in the way spatial dimensions (width and height) and the depth dimension are treated, making the network particularly adapted for multispectral images. The connectivity of the convolutional layer is local along the width and height, but the layer is fully connected along with depth.

A convolutional layer's parameters can also be seen as a set of spatially small-sized learnable filters or kernels. During the forward pass, the filters are convolved across the width and height dimensions of the input volume. This action produces a 2D activation map outputting the responses of the filter at each position of the input layer [26,32]. The output volume of a convolutional layer depends on three hyperparameters: the number of filters, the stride, and zero padding.

The number of filters in the same receptive field determines the depth of the output volume. A different filter activates for every different pattern. A set of units with the same receptive field is called the breadth of the output layer.

The stride used when the filters are slid along the spatial dimensions of the previous layer affects the height and width of the output volume. The higher the stride, the smaller is the output volume.

The input volume can be padded with zeros around the border to keep the information at the border. Without zero padding, the information carried by the pixels at the border of the input image vanishes quickly after successive convolutional layers. This artificially increases the size of the input layer, thereby increasing the size of the output layer.

Furthermore, the parameter-sharing scheme is used to reduce the number of parameters to be learned. It is based on the assumption that a useful feature at one position of the input layer is also useful at a different position. This means that the units on the same output depth slice use the same weights and biases. This explains the fact that the forward propagation through a convolutional layer is equivalent to convoluting a filter or kernel with the input layer.

## Pooling Layer

Typically, a pooling layer is inserted between the successive convolution layers. The pooling function replaces the output of a convolutional layer at a certain unit with the statistic of its neighboring units. The most popular pooling function used is the max-pooling method introduced by Zhou et al [33]. The pooling layer aims to make the system invariant to small input translations. This property gives more importance to whether a feature is present in the input rather than its exact position.

## CNN Feature Extraction and Classification

The combination of convolutional and pooling layers aims to learn the best features that can be extracted from the data set. This contrasts with most current methods that use handcrafted feature extraction techniques, such as those presented in the previous sections. These approaches can yield very good results but are usually sensitive to the data set and perform poorly when applied to different data sets. The combination of convolutional and pooling layers of a CNN provides a more versatile method for extracting features from images. The fully connected layers of the CNN correspond to the classifier. It aims at learning to classify learned features. As a result, a CNN is a unified versatile scheme for feature extraction and classification. As medical image classification is often a very complex task, it requires carefully manufactured feature sets for each type of data or even each different data set; doing just that with a unified framework, CNNs seem particularly adapted to the field.

## Data Set Description

The prostate gland and the colorectum have a similar tissue structure, with the tubular glandular mucosa—composed of epithelium and lamina propria—being their main functional tissue. This characteristic implies that these tissues are subject to development of the same types of tumors and cancers. Carcinomas are the most common type of malignant tumor and they are derived from epithelial cells [34]. Carcinomas are called adenocarcinomas when derived from glandular tissues, which is the case for both organs studied in this paper. All growths are not necessarily malignant, and benign polyps can occur [35]. They are usually noncancerous growths of the mucosa into the lumen and can be of different types.

Although most polyps are completely benign, such as hyperplastic polyps or hyperplasia, some types of polyps can transform into adenocarcinoma and can be considered as a precancerous stage. They are called adenomas and can be tubular or villous, depending on their growth patterns [36]. Hyperplastic polyps are characterized by an increase in the number of cells, resulting in an increased size of the tissue because of enhanced cell division. In contrast to an adenoma or a carcinoma, the division rate in a hyperplastic polyp returns to normal as soon as the stimulus is removed.

To best describe the different types of tumor recognized by pathologists, the following two data sets were used for the purpose of this study:

1. The prostate data set, which was used in previous works by Tahir and Bouridane [13] and Peyret et al [17], consists of 512 different multispectral prostate tumor tissue images of size 128×128. The images were taken at 16 spectral channels (500-650 nm) and 40× magnification power. The samples were evaluated by 2 highly experienced independent pathologists and labeled into four classes: 128 cases of stroma, which is normal muscular tissue, 128 cases of benign prostatic hyperplasia, a benign condition, 128 cases of prostatic intraepithelial neoplasia, a precancerous stage, and 128 cases of prostatic carcinoma, an abnormal tissue development corresponding to cancer.
2. The colorectal data set, which consists of multispectral colorectal histology data with a 40× magnification power, was developed by the University of Qatar in collaboration with Al-Ahli Hospital, Doha. It splits into 4 classes, each composed of 40 images. The images were acquired on a wider spectrum than the first data set, as it was spread on the visible and infrared ranges of the electromagnetic spectrum with an interval of 23 nm between each wavelength. That is to say, in the visible range, the wavelength interval is 23 nm starting from 465 to 695 nm, and in the infrared range, the wavelength interval is also 23 nm and ranges from 900 to 1590 nm. The special size was 128×60 pixels. The 4 classes were defined as carcinoma, containing images of cancerous colon biopsies; tubular adenoma, a precancerous stage; hyperplastic polyp, a benign polyp; and no remarkable pathology.

## Experiments

### Hardware and Software Specifications

To train deep CNNs, a GPU is required. The system used for this experiment was equipped with 1 NVIDIA K80 GPU and 4 central processing units. It had 61-GB RAM. Regarding software, Keras with a TensorFlow backend was used. Keras has the advantage of making available deep learning models alongside pretrained weights.

### Selected Architecture

The proposed CNN architecture evaluated for the task at hand was based on Visual Geometry Group 16 (VGG16) [37]. To design the proposed architecture, the last block of the convolutional layers of VGG16 was removed, and the number of filters per layer was halved. The idea is to reduce the capacity of the network because the interclass similarity in the data sets used for the task was high compared with the data set on which VGG16 was tested.

As represented in Figures 3 and 4, the overall proposed network architecture consists of a total of 13 layers with weights—the first 10 being convolutional layers, and the remaining 3 fully connected layers. The output of the last fully connected layer was fed to a SoftMax classifier, which is a generalization of the logistic regression classifier to the multiclass problem and produces a distribution of the 4 class labels. The network uses cross-entropy as a loss function.

Similar to VGG16, we decided to use a small kernel with a size of 3 pixels for every convolutional layer. The strategy of stacking convolutional layers with a small filter size is preferred to using a single large receptive field convolutional layer. For the same final receptive field, the former strategy includes nonlinearities (ReLU functions) at each layer, whereas the latter computes a simple linear function on the input, which makes the features less expressive. A stride of 1 was also adopted for the entire network to minimize information loss.

To achieve better control over the output size of each layer and maintain border information, a zero padding of 1 is added before each convolutional layer. The first 2 convolutional layers use 32 kernels followed by a 2 2 max-pooling layer. The max-pooling layer reduces the size of the output and thus the network capacity. The number of kernels is doubled in the next convolutional layer to compensate for this loss. Consequently, this sequence is followed by 2 convolutional layers with 64 filters, and then a new max-pooling layer is applied. This is followed by a series of 3 convolutional layers with 128 filters and a max-pooling layer. A final series of 3 convolutional layers with 256 filters and a max-pooling layer was applied. The neurons in the 3 fully connected layers with sizes of 1024, 1024, and 4, respectively, are connected to all neurons in the previous layer. The ReLU nonlinearity was applied to the output of every layer with weights.

Dropout is used after every max-pooling and fully connected layer to reduce overfitting. An early stopping strategy is also adopted to reduce the training time and regularization. Finally, data augmentation is carried out using the following transformations: each image is flipped along the 2 special axes, and 30 rotations in both directions are applied. This results in the generation of 27 fake images for each real data image. To ensure that the generalization is not overestimated, data set augmentation is performed after splitting the data set into training and test sets.

**Figure 3.** Illustration of the architecture of the proposed convolutional neural network for prostate cancer images. ReLU: rectified linear unit.

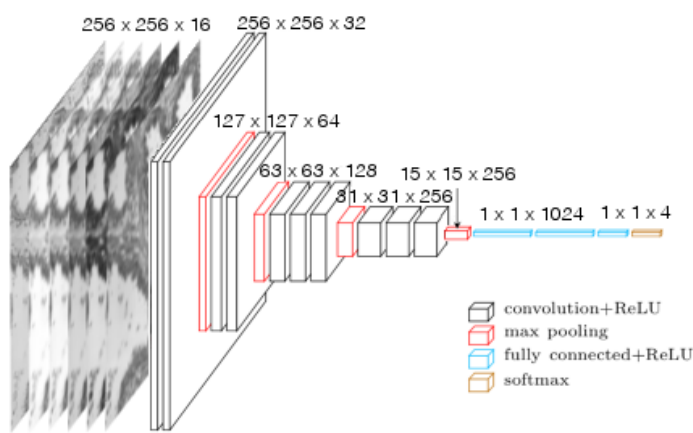

**Figure 4.** Illustration of the architecture of the proposed convolutional neural network for colorectal cancer images. ReLU: rectified linear unit.

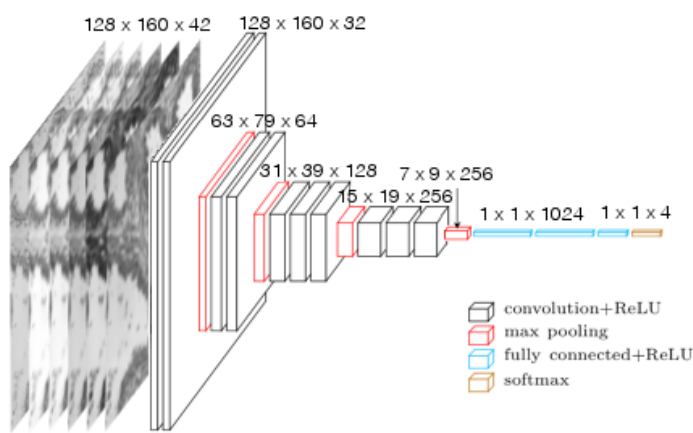

### *Details of Learning*

The weights of each layer are initialized using the Xavier initialization method [38], where the weights are drawn from a normal distribution centered on zero and with an SD of the following:

$$\sqrt{\frac{2}{N_{in}+N_{out}}}$$

where $N_{in}$ and $N_{out}$ are the numbers of input and output units, respectively. The network was trained separately on the 2 data sets.

The learning rate used was the same for all layers. It is optimized using a grid-search scheme, the results of which are presented in Figures 5 and 6. The learning rate selected for training was 0.0001 for both data sets.

For each model training, a 10-fold cross-validation technique was adopted to obtain a good estimate of the systems' generalization accuracy. This provides a large training set for better learning.

Figures 7 and 8 illustrate the evolution of the loss function during training for the prostate and colorectal data sets, respectively. Figures 9 and 10 show the evolution of their accuracies. It can be observed from these figures that the validation accuracy is very close to the training accuracy, which proves that the model is not in the overfitting regime. The higher variation in validation accuracy and loss can be explained by the smaller set used for validation compared with that used for training.

**Figure 5.** Validation accuracy obtained with different learning rates for the network trained on prostate data.

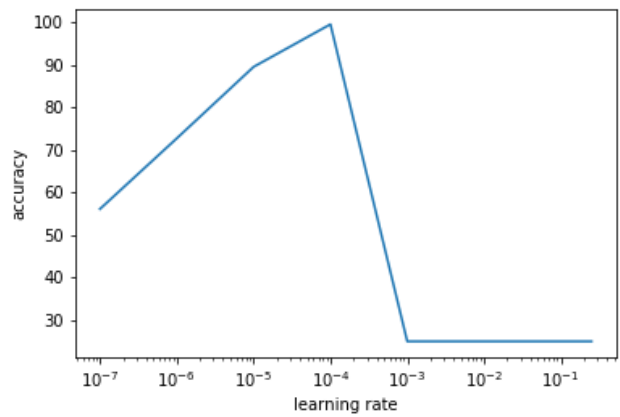

**Figure 6.** Validation accuracy obtained with different learning rates for the network trained on colorectal data.

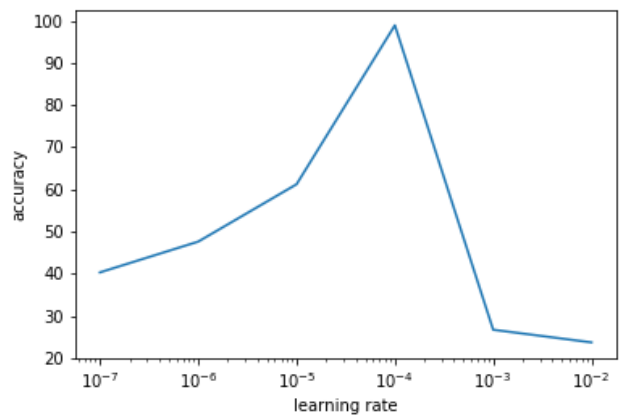

**Figure 7.** Loss function evolution during training for the prostate data set.

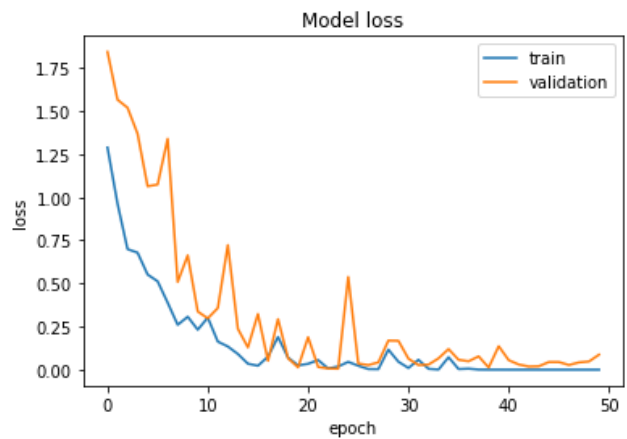

**Figure 8.** Loss function evolution during training for the colorectal data set.

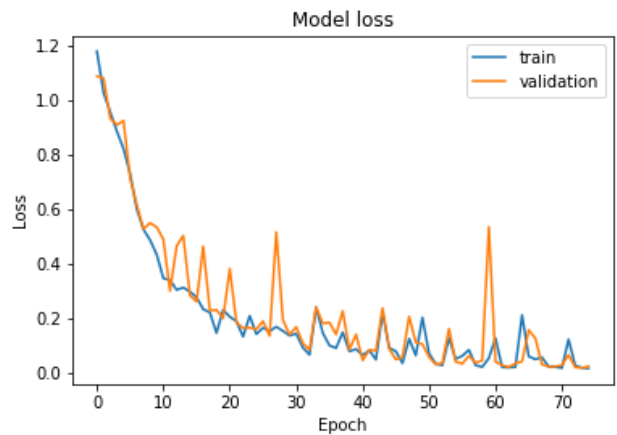

**Figure 9.** Accuracy evolution during training for the prostate data set.

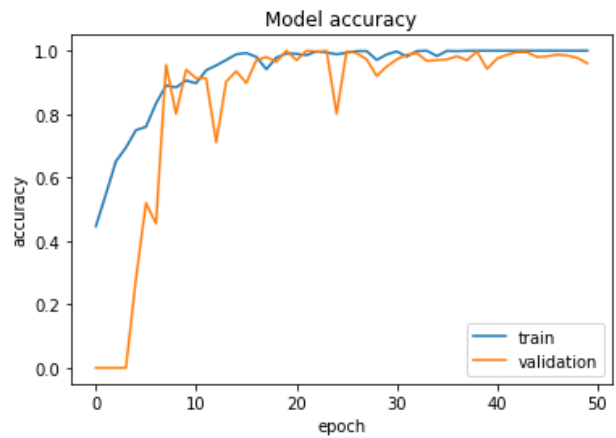

**Figure 10.** Accuracy evolution during training for the colorectal data set.

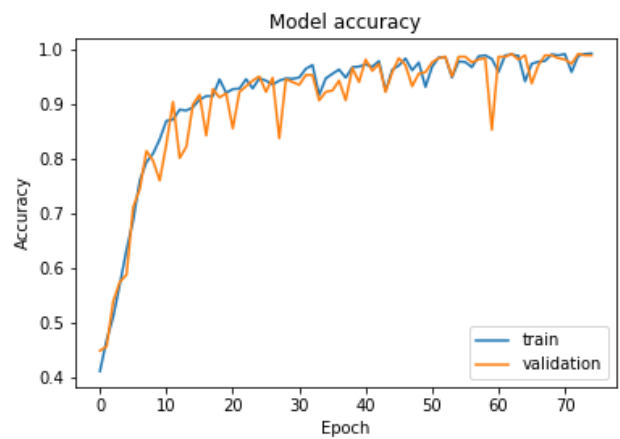

## Transfer Learning

Transfer learning consists of using a network previously trained on another data set to use the knowledge acquired during this learning task for the new task at hand [39]. In most transfer learning for image classification tasks, the ImageNet data set [40], which contains 1.2 million images with 1000 categories, is used for pretraining the network. When only a small data set is available, this allows the CNN to be trained on a very large data set and therefore train a high-capacity network that captures details without overfitting. Very deep networks also require a lot of time and very powerful machines equipped with multiple GPUs. Using pretrained networks can be advantageous when appropriate resources are not provided. Several transfer-learning scenarios are practical.

In the first scenario, the pretrained CNN is used as a fixed feature extractor. The convolutional layers of the network are kept with the weights determined during training on the ImageNet data set, and the pretrained fully connected layers are replaced with fully connected layers initialized with random weights. During training, only the newly added fully connected layers were marked as trainable. They used the features extracted by pretrained convolutional layers as inputs. These features are usually referred to as CNN codes [26,39].

Another strategy is to retrain the fully connected layers from scratch to fine-tune the weights of the pretrained convolutional layers by continuing back-propagation. Either all the convolutional layers can be retuned or only some of the higher-level layers to avoid overfitting. This derives from the observation that the lower-level layers usually learn more generic features, such as edge detectors, that can be used for many different learning tasks. In contrast, the high-level layers tend to learn features that are more specific to the characteristics of the classes of the original data set.

In this study, only the first scenario was investigated. The pretrained CNNs are very deep and require very high computational power to be retuned. Using them as feature extractors is, in fact, equivalent to training only a relatively shallow MLP. The proposed architecture was compared with popular CNN architectures: VGG16 [37], InceptionV3 [21], and ResNet50 [22]. These networks were initialized with the weights obtained when pretraining them on the ImageNet data set. However, InceptionV3 and ResNet50 are very deep networks (48 and 152 layers, respectively), and a minimum input image size is required. InceptionV3 requires a minimum width and height of 139 pixels and ResNet50 of 197 pixels. The images of the colorectal data set were smaller, and zero padding was added to reach the required dimensions. Moreover, the ImageNet images are RGB images and therefore have a depth of 3 channels. To meet the dimension requirements, a principal component analysis (PCA) was carried out to reduce the dimensionality of the multiscale images to 3 channels.

## Results and Discussion

### Principal Results and Findings

To visualize the effect of the kernels on images through the network, Figures 11 and 12 present examples of outputs of the first convolutional layer of the networks trained with the prostate and colorectal data sets, respectively. Figures 13 and 14 depict examples of outputs of the last convolutional layers of the same networks. It can be observed that after the first layer, the outputs are very similar to the input image, for instance, with transformations resembling edge detections. Once the image has its own through the network, different regions or features of the input image are represented in the outputs of the last convolutional layer. Thus, the different layers learn a succession of transformations, leading to an isolation of relevant regions or features of the input image. The fully connected layers of the network are then able to classify these particular features into the 4 classes.

**Figure 11.** Example of an output of the first convolutional layer for the network trained on the prostate data set.

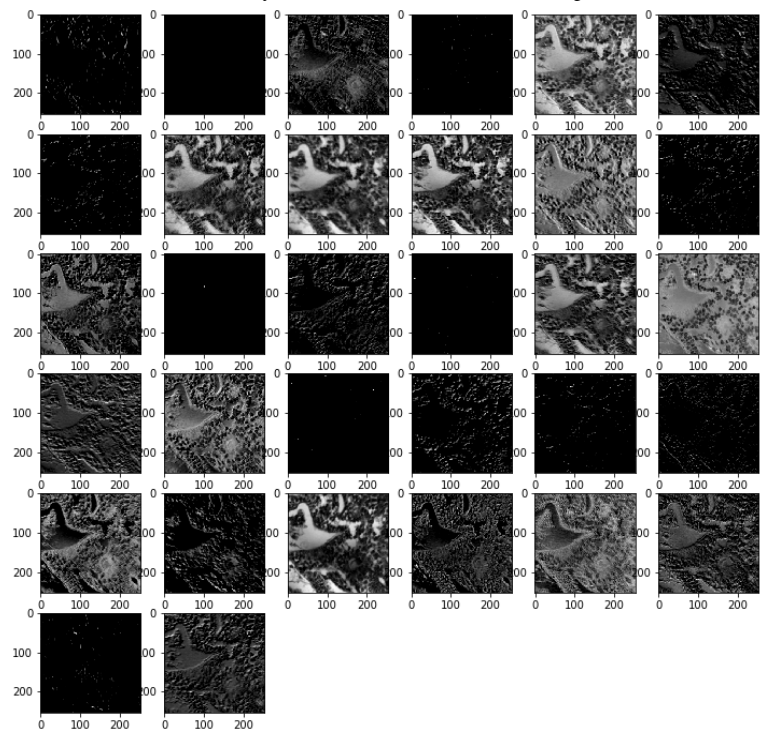

**Figure 12.** Example of an output of the first convolutional layer for the network trained on the colorectal data set.

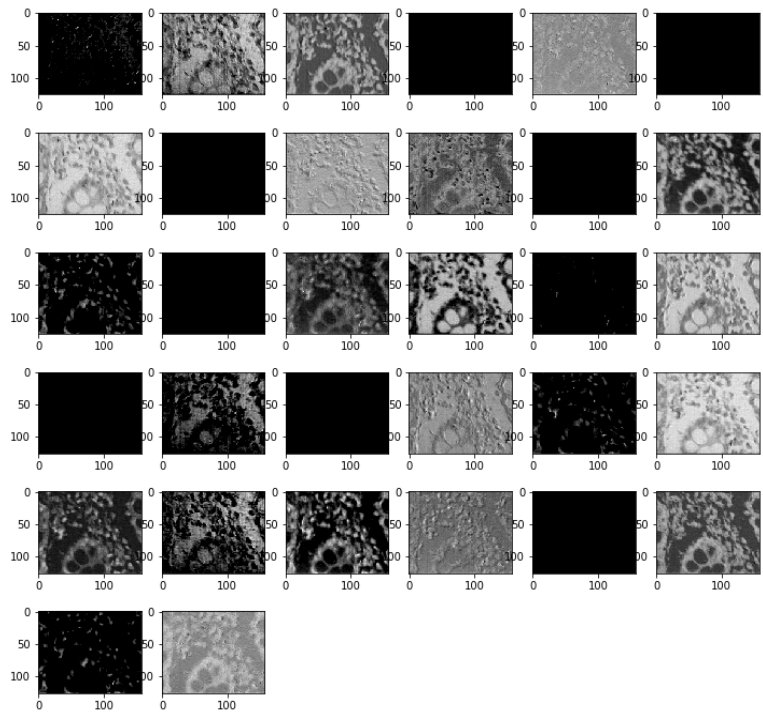

**Figure 13.** Example of an output of the last convolutional layer for the network trained on the prostate data set.

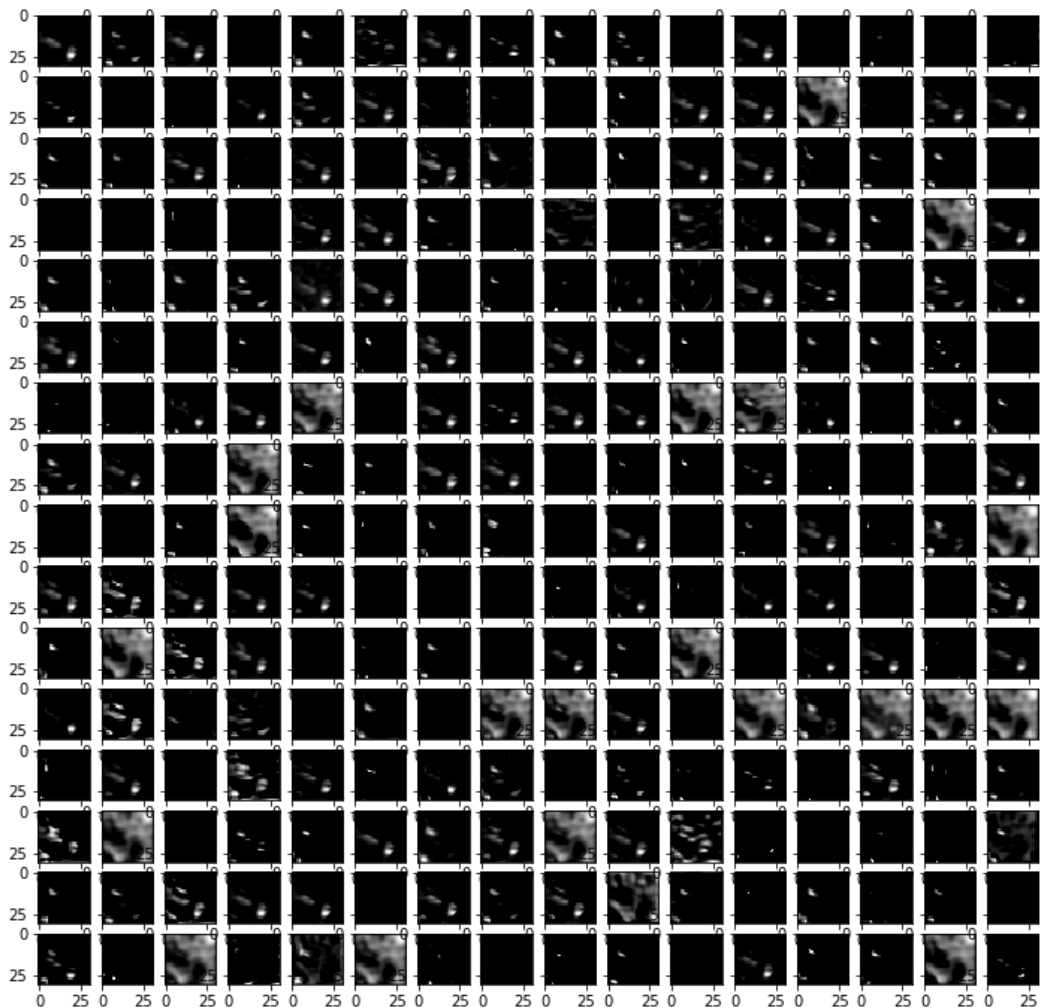

**Figure 14.** Example of an output of the last convolutional layer for the network trained on the colorectal data set.

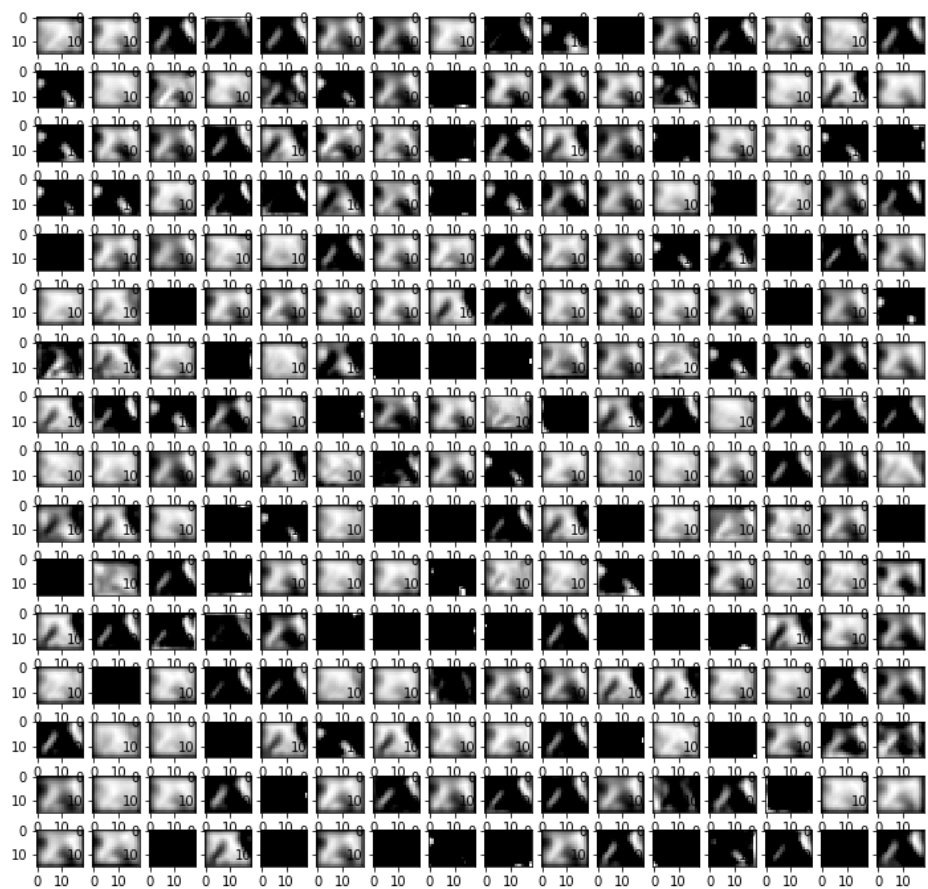

Table 1 displays the validation and test accuracies obtained using the prostate and colorectal data sets for different CNN models. This shows that the validation and test accuracies are very close, proving a good generalization of the systems and that overfitting was avoided.

The proposed CNN model achieved an average test accuracy of 99.8% and 99.5% for the prostate and colorectal data sets, respectively. Table 2 shows that the optimal CNN weights were obtained after 44 and 70 epochs, respectively. The VGG16 model initialized with Xavier weights trains very quickly for the prostate data set; the optimal validation accuracy was obtained after 19 epochs, as illustrated in Table 2. However, it is less efficient at learning for the colorectal data set and requires as many as 70 epochs to obtain the minimum validation loss. The results also show slight overfitting for the colorectal data set, as the validation accuracy is lower than the training accuracy. This is because of the high capacity of the network. When using this network with pretrained weights from ImageNet, the training loss reaches a minimum after only a few epochs, but the validation loss shows that the network overfits marginally for both data sets. The test accuracy was also lower than that of the proposed CNN by 99.5% and 98.1%, respectively. This is because the CNN codes learned with the ImageNet data set are not as adapted to the classification task at hand as those learned with the proposed CNN. The InceptionV3 model shows a higher overfitting and a lower generalization for both data sets with 99.0% and 94.5% accuracy for the prostate and colorectal data sets, respectively. This shows once again that the CNN codes learned on the ImageNet data set with this network are not adapted to the classification task at hand. Finally, the pretrained ResNet50 achieved optimal accuracy with the lowest number of epochs: 5 and 22 for the prostate and colorectal data sets, respectively. It also achieves 100% average accuracy for the prostate data set, outperforming the proposed CNN, and 99% for the colorectal data set, which is slightly lower than the proposed data set. This lower performance compared with the proposed CNN architecture for the colorectal data set might be owing to some loss of information when performing PCA on the 42 channels of the colorectal data set images. The prostate data set consisted of images with only 16 channels, and it is logical that the loss of information is not as important during this transformation.

Therefore, the proposed CNN architecture is more adapted to the task at hand than the other methods it was compared with. However, ResNet50 shows very good performance when used as a feature extractor and is trained with fewer epochs. In every case, it was noted that the colorectal data set is more prone to overfitting. This is probably owing to the size of the images, which are spatially smaller than those for the prostate data set. Therefore, a model with the correct capacity for the prostate data set might be overestimated for the colorectal data set.

Table 1. Validation and test accuracy comparison of different architectures.[a]

| Method | Prostate data set (% accuracy), mean (SD) | | Colorectal data set (% accuracy), mean (SD) | |
| --- | --- | --- | --- | --- |
| | Validation | Test | Validation | Test |
| Proposed CNN[b] | 100 | 99.8 (0.1) | 100 | 99.5 (0.1) |
| VGG16[c] Xavier initial | 100 | 99.6 (0.1) | 99.0 (0.1) | 99.2 (0.1) |
| VGG16 pretrained | 100 | 99.5 (0.1) | 97.5 (0.2) | 98.1 (0.1) |
| InceptionV3 pretrain | 98.8 (0.2) | 99.0 (0.1) | 92.3 (0.3) | 94.5 (0.3) |
| ResNet50 pretrained | 100 | 100 | 99.5 (0.1) | 99.0 (0.2) |

[a]SD values have been provided wherever applicable.
[b]CNN: convolutional neural network.
[c]VGG16: Visual Geometry Group 16.

Table 2. Number of epochs until early stopping.

| Method | Prostate data set | Colorectal data set |
| --- | --- | --- |
| Proposed CNN[a] | 44 | 70 |
| VGG16[b] Xavier initialization | 19 | 70 |
| VGG16 pretrained | 10 | 38 |
| InceptionV3 pretrained | 48 | 53 |
| ResNet50 pretrained | 5 | 22 |

[a]CNN: convolutional neural network.
[b]VGG16: Visual Geometry Group 16.

## Comparison Against Other Machine Learning Methods

Table 3 shows the test accuracy of the best-performing CNN architectures compared with other methods from Tahir et al [15], Bouatemane et al [16], Haj-Hassan et al [25], and Peyret et al [17] stacked multispectral multiscale local binary pattern (MMLBP) + gray-level co-occurrence matrix (GLCM), and concatenated local binary pattern [18]. Regarding the prostate data set, 5 systems have an accuracy above 99%: Bouatemane et al [16], Stacked MMLBP+GLCM, the proposed CNN, Haj-Hassan et al [25], and ResNet50 with pretrained weights. The highest classification accuracy was achieved using ResNet50 with 100% accuracy. The proposed CNN and the study by Bouatemane et al [16] achieved 99.8% accuracy; however, the SD was not given for the latter. Therefore, it is not possible to determine the precision of the accuracy estimation. The stacked MMLBP+GLCM system achieves 99.5% (SD 0.3 pp), which makes this performance similar to that of the proposed CNN. However, a higher SD shows lower precision in the accuracy estimation. Therefore, the proposed CNN was preferred. The study by Haj-Hassan et al [25] achieved a 99.17% accuracy with segmentation. Their system without this preprocessing phase achieved an accuracy of 79.23%.

This can be explained by the lower capacity of their model compared with ours. This has the advantage of requiring less processing power. However, this is counterbalanced by the fact that their system requires a preprocessing phase with the intervention of a pathologist, which dramatically increases the processing time of the system. Furthermore, they state that their CNN model requires 500 epochs to be trained, which is much higher than that of the proposed model. With respect to the colorectal data set, Peyret et al [17] stacked MMLBP+GLCM system and the proposed CNN both provided the same accuracy and SD. They outperform ResNet50 with pretrained weights by 0.5 pp.

Finally, when considering the results obtained with both data sets, the stacked MMLBP+GLCM system and the proposed CNN appear to provide the most stable results as well as the highest accuracy. However, on average, the SD of the accuracy achieved by the proposed CNN is lower than that obtained with the stacked MMLBP+GLCM system. The performance of the ResNet50 network seems to be more dependent on the data set used. Moreover, it would be interesting to compare the system proposed by Bouatemane et al [16] using the colorectal data set to verify whether it performs as well on different data sets. Considering the current information available on the system performance and with the data sets available, the proposed CNN is selected as the best-performing system in terms of accuracy for the classification task at hand.

**Table 3.** Accuracy comparison against other methods.[a]

| Method | Prostate data set (% accuracy), mean (SD) | Colorectal data set (% accuracy), mean (SD) |
| --- | --- | --- |
| Tahir et al [15] | • 98.9 | • N/A[b] |
| Bouatemane et al [16] | • 99.83 | • N/A |
| Concatenated LBP[c] [18] | • 92.4 (0.4)<br>• 99.5 (0.3) | • 88.2 (0.5)<br>• 99.5 (0.1) |
| Stacked MMLBP[d] + GLCM[e] [41] | • 99.5 (0.3) | • 99.5 (0.1) |
| Haj-Hassan et al [25] | • 99.17 | • N/A |
| Proposed CNN | • 99.8 (0.1) | • 99.5 (0.1) |
| ResNet50 pretrained | • 100 | • 99.0 (0.2) |

[a]SD values have been provided wherever applicable.

[b]N/A: not applicable.

[c]LBP: local binary pattern.

[d]MMLBP: multispectral multiscale local binary pattern.

[e]GLCM: gray-level co-occurrence matrix.

## Computational Complexity Analysis

In computer-aided diagnosis systems (CADSs), an unlabeled image is fed to a previously trained system. Consequently, the time used to process this image is decisive, as it is crucial that the CADS works on the web. However, the forward pass of an image through the CNN architectures studied in this study is computationally nonexpensive. Table 4 displays the classification times per image for all CNN architectures tested. This demonstrates that only a few milliseconds are required to classify one image once the CNN has been trained. However, it must be noted that the proposed CNN architecture is much quicker at classifying images than the others. This is because, for the architectures described in the literature and the pretrained networks, a PCA must be carried out to reduce to 3 the number of channels of the image to be classified. This preprocessing stage lengthens the total classification time.

As mentioned, training is performed only once when a CADS is created. Consequently, training time is not a critical measure of the problem at hand. However, the computational complexity of deep learning systems can rapidly increase significantly. Such architectures require high-performing hardware, including GPUs. Some extremely deep architectures can also entail several weeks of training time [26]. Such long training times considerably slowed down the CADS development process. To verify that the proposed system can be trained within a reasonable duration, a comparison of the training times for each architecture was carried out (Table 5). The computational times depending on the hardware and software used, it is not possible to compare the CNN architectures with other classification systems proposed in other published works. However, this is one of the first attempts to use deep learning for this application. Therefore, this section aims to establish the ability of deep learning systems to be trained in a short period using the data sets used.

Unsurprisingly, Table 5 demonstrates that pretrained networks have a much shorter training time per epoch owing to the reduced number of layers to be trained; ResNet50 and InceptionV3 can be trained in a few minutes. When considering this measure of performance, the best architecture was ResNet50. However, the total training time for every CNN model is <2 hours, making it a reasonable time for developing a CADS.

**Table 4.** Average convolutional neural network (CNN) classification computation times for 1 image.

| Method | Prostate data set (ms) | Colorectal data set (ms) |
| --- | --- | --- |
| Proposed CNN | 14 | 7 |
| VGG16[a] Xavier initial | 75 | 42 |
| VGG16 pretrained | 75 | 42 |
| InceptionV3 pretrained | 63 | 42 |
| ResNet50 pretrained | 65 | 47 |

[a]VGG16: Visual Geometry Group 16.

Table 5. Average convolutional neural network (CNN) training computation times for the complete data set.

| Method | Prostate data set (seconds) | | Colorectal data set (seconds) | |
| --- | --- | --- | --- | --- |
| | Time per epoch | Total training | Time per epoch | Total training |
| Proposed CNN | 90 | 3780 | 45 | 2925 |
| VGG16[a] Xavier initial | 245 | 4655 | 97 | 6790 |
| VGG16 pretrained | 83 | 3154 | 35 | 1400 |
| InceptionV3 pretrained | 39 | 1755 | 15 | 705 |
| ResNet50 pretrained | 41 | 205 | 32 | 704 |

[a]VGG16: Visual Geometry Group 16.

## Conclusions

In this paper, the proposed CNN architecture was detailed and compared with previously trained network models used as feature extractors. These CNNs were also compared with other classification methods from other published studies. The proposed CNN demonstrated excellent performance compared with pretrained CNNs and other classification methods. The computational complexity of the CNNs was also analyzed, and it was demonstrated that the proposed CNN is faster at classifying images than pretrained networks because it avoids a preprocessing phase. The conclusion of this overall analysis is that the proposed CNN architecture was globally the best-performing system for classifying colorectal and prostate tumor images.

## Acknowledgments

This research project was supported by a grant from the Research Supporting Program (Project Number: RSP2022R281), King Saud University, Riyadh, Saudi Arabia.

## Conflicts of Interest

None declared.

## References

1. Ferlay J, Soerjomataram I, Dikshit R, Eser S, Mathers C, Rebelo M, et al. Cancer incidence and mortality worldwide: sources, methods and major patterns in GLOBOCAN 2012. Int J Cancer 2015 Mar 01;136(5):E359-E386 [FREE Full text] [doi: 10.1002/ijc.29210] [Medline: 25220842]
2. Heidenreich A, Bellmunt J, Bolla M, Joniau S, Mason M, Matveev V, European Association of Urology. EAU guidelines on prostate cancer. Part 1: screening, diagnosis, and treatment of clinically localised disease. Eur Urol 2011 Jan;59(1):61-71. [doi: 10.1016/j.eururo.2010.10.039] [Medline: 21056534]
3. Humphrey P. Prostate Pathology. Chicago: American Society for Clinical Pathology; 2003.
4. Thomas GD, Dixon MF, Smeeton NC, Williams NS. Observer variation in the histological grading of rectal carcinoma. J Clin Pathol 1983 Apr;36(4):385-391 [FREE Full text] [doi: 10.1136/jcp.36.4.385] [Medline: 6833507]
5. Kronz JD, Westra WH, Epstein JI. Mandatory second opinion surgical pathology at a large referral hospital. Cancer 1999 Dec 01;86(11):2426-2435. [Medline: 10590387]
6. Tobore I, Li J, Yuhang L, Al-Handarish Y, Kandwal A, Nie Z, et al. Deep learning intervention for health care challenges: some biomedical domain considerations. JMIR Mhealth Uhealth 2019 Aug 02;7(8):e11966 [FREE Full text] [doi: 10.2196/11966] [Medline: 31376272]
7. Owais M, Arsalan M, Mahmood T, Kang JK, Park KR. Automated diagnosis of various gastrointestinal lesions using a deep learning-based classification and retrieval framework with a large endoscopic database: model development and validation. J Med Internet Res 2020 Nov 26;22(11):e18563 [FREE Full text] [doi: 10.2196/18563] [Medline: 33242010]
8. Zhao Z, Wu C, Zhang S, He F, Liu F, Wang B, et al. A novel convolutional neural network for the diagnosis and classification of rosacea: usability study. JMIR Med Inform 2021 Mar 15;9(3):e23415 [FREE Full text] [doi: 10.2196/23415] [Medline: 33720027]
9. Mosquera-Lopez C, Agaian S, Velez-Hoyos A, Thompson I. Computer-aided prostate cancer diagnosis from digitized histopathology: a review on texture-based systems. IEEE Rev Biomed Eng 2015;8:98-113. [doi: 10.1109/RBME.2014.2340401] [Medline: 25055385]
10. Kunhoth S, Al Maadeed S. Multispectral biopsy image based colorectal tumor grader. In: Valdés Hernández M, González-Castro V, editors. Medical Image Understanding and Analysis. Cham: Springer; 2017:330-341.
11. Roula M, Diamond J, Bouridane A, Miller P, Amira A. A multispectral computer vision system for automatic grading of prostatic neoplasia. In: Proceedings IEEE International Symposium on Biomedical Imaging. 2002 Presented at: Proceedings

    IEEE International Symposium on Biomedical Imaging; Jul 7-10, 2002; Washington, DC, USA. [doi: 10.1109/ISBI.2002.1029226]
12. Roula MA. Machine vision and texture analysis for the automated identification of tissue pattern in prostatic neoplasia. PhD Thesis. Belfast, Northern Ireland: Queen's University of Belfast; 2004.
13. Tahir M, Bouridane A. Novel round-robin tabu search algorithm for prostate cancer classification and diagnosis using multispectral imagery. IEEE Trans Inf Technol Biomed 2006 Oct;10(4):782-793. [doi: 10.1109/titb.2006.879596] [Medline: 17044412]
14. Tahir M, Bouridane A, Kurugollu F. Simultaneous feature selection and feature weighting using Hybrid Tabu Search/K-nearest neighbor classifier. Pattern Recog Letters 2007 Mar;28(4):438-446 [FREE Full text] [doi: 10.1016/j.patrec.2006.08.016]
15. Tahir M, Bouridane A, Roula M. Prostate cancer classification using multispectral imagery and metaheuristics. In: Computational Intelligence in Medical Imaging. London, United Kingdom: Chapman and Hall; 2009.
16. Bouatmane S, Roula M, Bouridane A, Al-Maadeed S. Round-Robin sequential forward selection algorithm for prostate cancer classification and diagnosis using multispectral imagery. Mach Vision Apps 2010 Sep 16;22(5):865-878. [doi: 10.1007/s00138-010-0292-x]
17. Peyret R, Khelifi F, Bouridane A, Al-Maadeed S. Automatic diagnosis of prostate cancer using multispectral based linear binary pattern bagged codebooks. In: Proceedings of the 2nd International Conference on Bio-engineering for Smart Technologies (BioSMART). 2017 Presented at: 2nd International Conference on Bio-engineering for Smart Technologies (BioSMART); Aug 30 -Sep 1, 2017; Paris, France. [doi: 10.1109/biosmart.2017.8095322]
18. Peyret R, Bouridane A, Al-Maadeed S, Kunhoth S, Khelifi F. Texture analysis for colorectal tumour biopsies using multispectral imagery. In: Proceedings of the 37th Annual International Conference of the IEEE Engineering in Medicine and Biology Society (EMBC). 2015 Presented at: 37th Annual International Conference of the IEEE Engineering in Medicine and Biology Society (EMBC); Aug 25-29, 2015; Milan, Italy. [doi: 10.1109/embc.2015.7320057]
19. Lasch P, Chiriboga L, Yee H, Diem M. Infrared spectroscopy of human cells and tissue: detection of disease. Technol Cancer Res Treat 2002 Feb;1(1):1-7 [FREE Full text] [doi: 10.1177/153303460200100101] [Medline: 12614171]
20. Krizhevsky A, Sutskever I, Hinton GE. ImageNet classification with deep convolutional neural networks. Commun ACM 2017 May 24;60(6):84-90 [FREE Full text] [doi: 10.1145/3065386]
21. Szegedy C, Liu W, Jia Y, Sermanet P, Reed S, Anguelov D, et al. Going deeper with convolutions. In: Proceedings of the IEEE Conference on Computer Vision and Pattern Recognition (CVPR). 2015 Presented at: IEEE Conference on Computer Vision and Pattern Recognition (CVPR); Jun 7-12, 2015; Boston, MA URL: http://arxiv.org/abs/1409.4842 [doi: 10.1109/cvpr.2015.7298594]
22. He K, Zhang X, Ren S, Sun J. Deep residual learning for image recognition. arXiv. Preprint posted online December 10, 2015 [FREE Full text]
23. Silver D, Huang A, Maddison CJ, Guez A, Sifre L, van den Driessche G, et al. Mastering the game of Go with deep neural networks and tree search. Nature 2016 Jan 28;529(7587):484-489. [doi: 10.1038/nature16961] [Medline: 26819042]
24. Silver D, Schrittwieser J, Simonyan K, Antonoglou I, Huang A, Guez A, et al. Mastering the game of Go without human knowledge. Nature 2017 Oct 18;550(7676):354-359. [doi: 10.1038/nature24270] [Medline: 29052630]
25. Haj-Hassan H, Chaddad A, Harkouss Y, Desrosiers C, Toews M, Tanougast C. Classifications of multispectral colorectal cancer tissues using convolution neural network. J Pathol Inform 2017;8:1 [FREE Full text] [doi: 10.4103/jpi.jpi_47_16] [Medline: 28400990]
26. Goodfellow I, Bengio Y, Courville A. Deep Learning. Cambridge, Massachusetts, United States: MIT Press; 2016.
27. Lecun Y, Bottou L, Bengio Y, Haffner P. Gradient-based learning applied to document recognition. Proc IEEE 1998 Nov;86(11):2278-2324. [doi: 10.1109/5.726791]
28. Rumelhart D, Durbin R, Golden R, Chauvin Y. Backpropagation: the basic theory. In: Backpropagation Theory, Architectures, and Applications. Mahwah: Lawrence Erlbaum Associates; 1995.
29. Rumelhart DE, Hinton GE, Williams RJ. Learning representations by back-propagating errors. Nature 1986 Oct;323(6088):533-536. [doi: 10.1038/323533a0]
30. LeCun Y, Bottou L, Orr G, Müller K. Efficient BackProp. In: Neural Networks: Tricks of the Trade. Berlin, Heidelberg: Springer; 2012.
31. LeCun Y, Boser B, Denker JS, Henderson D, Howard RE, Hubbard W, et al. Backpropagation applied to handwritten zip code recognition. Neural Comput 1989 Dec;1(4):541-551. [doi: 10.1162/neco.1989.1.4.541]
32. LeCun Y, Bengio Y, Hinton G. Deep learning. Nature 2015 May 28;521(7553):436-444. [doi: 10.1038/nature14539] [Medline: 26017442]
33. Zhou Y, Chellappa R, Vaid A, Jenkins B. Image restoration using a neural network. IEEE Trans Acoust Speech Signal Processing 1988 Jul;36(7):1141-1151. [doi: 10.1109/29.1641]
34. Lackie J. A Dictionary of Biomedicine. Oxford, UK: Oxford University Press; 2010.
35. Jass J, Sobin L. Histological classification of intestinal tumours. In: Histological Typing of Intestinal Tumours. Berlin, Heidelberg: Springer; 1989.

36. Understanding your pathology report: colon polyps (sessile or traditional serrated adenomas). American Cancer Society. URL: https://www.cancer.org/treatment/understanding-your-diagnosis/tests/understanding-your-pathology-report/colon-pathology/colon-polyps-sessile-or-traditional-serrated-adenomas.html [accessed 2022-01-12]
37. Simonyan K, Zisserman A. Very deep convolutional networks for large-scale image recognition. arXiv. Preprint posted online September 4, 2014 [FREE Full text]
38. Glorot X, Bengio Y. Understanding the difficulty of training deep feedforward neural networks. In: Proceedings of the Thirteenth International Conference on Artificial Intelligence and Statistics. 2010 Presented at: Proceedings of the Thirteenth International Conference on Artificial Intelligence and Statistics; May 13-15, 2010; Sardinia, Italy.
39. Transfer learning. CiteSeerX. URL: http://citeseerx.ist.psu.edu/viewdoc/summary?doi=10.1.1.146.1515 [accessed 2022-01-12]
40. Russakovsky O, Deng J, Su H, Krause J, Satheesh S, Ma S, et al. ImageNet large scale visual recognition challenge. Int J Comput Vis 2015 Apr 11;115(3):211-252. [doi: 10.1007/s11263-015-0816-y]
41. Peyret R, Bouridane A, Khelifi F, Tahir M, Al-Maadeed S. Automatic classification of colorectal and prostatic histologic tumor images using multiscale multispectral local binary pattern texture features and stacked generalization. Neurocomputing 2018 Jan;275(C):83-93 [FREE Full text] [doi: 10.1016/j.neucom.2017.05.010]

## Abbreviations

**CADS:** computer-aided diagnosis system
**CNN:** convolutional neural network
**GLCM:** gray-level co-occurrence matrix
**GPU:** graphic processing unit
**MLP:** multilayer perceptron
**MMLBP:** multispectral multiscale local binary pattern
**PCA:** principal component analysis
**ReLU:** rectified linear unit
**RGB:** red, green, blue
**VGG16:** Visual Geometry Group 16